\documentclass[aps,pre,showpacs,floatfix,superscriptaddress,preprint]{revtex4}
\usepackage{graphicx}
\usepackage{amsmath}
\usepackage{amsfonts}
\usepackage{epstopdf}

\newcommand{\deff}{{\cal D}_{\!e\!f\!\!f}}
\newcommand{\defc}{{\cal D}_{c}}
\newcommand{\norm}{{\cal N}}
\newcommand{\sinc}{\mbox {sinc}}

\begin{document}

\title{Mobility induces global synchronization of oscillators in periodic extended systems}

\author{Fernando Peruani} 
\affiliation{CEA-Service de Physique de l'Etat Condens\'{e}, Centre d'Etudes de Saclay, 91191 Gif-sur-Yvette, France}
\affiliation{Max Planck Institute for the Physics of Complex Systems, N\"othnitzer Str. 38, 01187 Dresden, Germany}
\author{Ernesto M. Nicola}
\affiliation{Max Planck Institute for the Physics of Complex Systems, N\"othnitzer Str. 38, 01187 Dresden, Germany}
\affiliation{IFISC, Institute for Cross-Disciplinary Physics and Complex Systems (CSIC-UIB), Campus Universitat Illes Balears, E-07122 Palma de Mallorca, Spain}
\author{Luis G. Morelli}  
\affiliation{Max Planck Institute for the Physics of Complex Systems, N\"othnitzer Str. 38, 01187 Dresden, Germany}
\affiliation{Departamento de F\'isica, FCEyN, UBA, Ciudad Universitaria,~1428 Buenos Aires,~Argentina}\affiliation{Max~Planck~Institute~of~Molecular~Cell~Biology~and~Genetics,~Pfotenhauerstr.~108,~01307~Dresden,~Germany}

\date{\today}

\begin{abstract}
We study synchronization of locally coupled noisy phase oscillators that move diffusively in a one-dimensional ring.
Together with the disordered and the globally synchronized states, the system also exhibits wave-like states displaying local order.
We use a statistical description valid for a large number of oscillators to show that for any finite system there is a critical mobility above which all wave-like solutions become unstable.
Through Langevin simulations, we show that the transition to global synchronization is mediated by a shift in the relative size of attractor basins associated to wave-like states.
Mobility disrupts these states and paves the way for the system to attain global synchronization.
\end{abstract}

\maketitle

Synchronization of oscillators is a widespread phenomenon in nature~\cite{manrubia,pikovsky,sync}.
In biology, synchronization can occur at scales that range from groups of single cells to ensembles of complex organisms~\cite{winfree}.
When oscillators hold fixed positions in space and the interaction that drives synchronization is short ranged, spatial and temporal patterns can self-organize.
Such is the case in cardiac tissue, where cells generate spiral patterns that shape the heartbeats~\cite{mackey}.
Also in central pattern generators, the oscillating neural network self-organizes to produce coordinated movements of the body~\cite{cohen82}.

A different situation arises when the oscillators are not fixed in space but are able to move around.
The problem of synchronization of moving oscillators has many applications in the domain of chemistry~\cite{taylor09}, biology~\cite{demonte07}, and technology~\cite{buscarino06}.
Small porous particles loaded with the catalyst of the Belousov-Zhabotinsky reaction behave as individual chemical oscillators,
undergoing a density-dependent synchronization transition as the stirring rate is increased~\cite{taylor09}.
The same particles support wave propagation in the form of dynamic target and spiral patterns when the particles are not moving~\cite{tinsley09}.
This phenomenon illustrates a wider scenario: mobility and mixing remove local defects and patterns, enabling global order.
This effect has far reaching consequences in finite systems.
For example, in ecosystems of competing populations with cyclic interactions,
biodiversity can be sustained if dispersal is local, but it is lost when dispersal occurs over large length scales~\cite{kerr02}.
The dynamics of such cyclic competition was described by a complex Ginzburg-Landau equation near a Hopf bifurcation,
displaying complex oscillatory patterns indicative of biodiversity for low mobility, 
while in the case of high mobility diversity is wiped out~\cite{reichenbach07a,reichenbach07b}.

\begin{figure}[b]
\centering\resizebox{13.5cm}{!}{\rotatebox{0}{\includegraphics{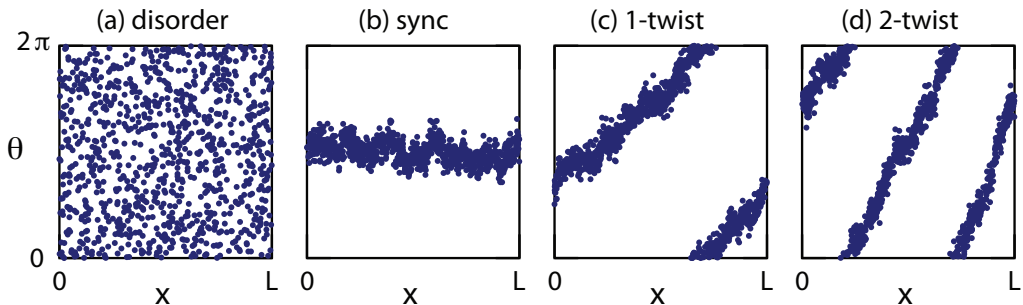}}}
\caption{The system, Eqs.~(\ref{eq:osc_i_theta})-(\ref{eq:osc_i_x}), exhibits different kinds of states: (a) disorder,
(b) partial synchronization, and twisted states: \emph{e.g.} (c) 1-twist and (d) 2-twist.
Each dot $(x,\theta)$ represents an oscillator.
Parameters in arbitrary units: $N=1000$, $L=2\pi$, $r=L/100$, $\omega=0$, $\gamma=0.1$ and $\sqrt{2D}=0.01$.
In (a) $\sqrt{2C}=0.35$ and in (b)-(d) $\sqrt{2C}=0.10$. States (b)-(d) coexist. 
The snapshots were taken after $15000$ time units, starting from random initial conditions.} \label{fig:snapshots}
\end{figure}

In this paper we study the effects of mobility ---spatial diffusion--- on the macroscopic collective dynamics
of locally coupled, moving phase oscillators subjected to noise, in a one-dimensional ring.
When oscillators are fixed in space, these systems can exhibit a series of
steady states where local order is present~\cite{shiogai03,kuramoto06,wiley06}.
Such states have been called $m$-twist solutions~\cite{wiley06}, Fig.~\ref{fig:snapshots}.
Here we show that mobility can destabilize all $m$-twist solutions, enhancing the stability of the synchronized solution.
We find that in finite systems there is a critical mobility above which either the synchronized or the disordered state is stable.

\section{Diffusing phase oscillators} 
We consider an ensemble of $N$ identical phase oscillators that diffuse on a ring of perimeter $L$.
Oscillators are coupled to other oscillators in their local
neighborhood, within an interaction range $r$.
The dynamics of phase and position is described by
\begin{eqnarray}\label{eq:osc_i_theta}
\dot{\theta}_i(t)       &=& \omega - \gamma \left[ \frac{1}{n_i}  \!\!\!\!\!\! \sum_{\quad |x_i-x_j|<r} \!\!\!\!\!\!\! \sin(\theta_i - \theta_j) \right] + \sqrt{2 C} \,\, \xi_{\theta,i}(t)  \\
\label{eq:osc_i_x} \dot{x}_i(t) &=& \sqrt{2 D} \,\, \xi_{x,i}(t)  \, ,
\end{eqnarray}
where $i=1,\ldots,N$ is the oscillator label, $\theta_i(t)$ and
$x_i(t)$ are the phase and position of the $i$-th oscillator at time
$t$, $\omega$ is the autonomous frequency, 
and $\gamma$ is the coupling strength ---whose inverse characterizes the typical relaxation time of the interaction.
Each oscillator interacts with its $n_i$ neighbors in the range $r$ through the coupling function in brackets, 
which defines an attractive interaction towards the local average of the phase.
In steady state the spatial density is uniform and the number of neighbors is on average constant, $n=n_i=N \, r/L$.
With this definition of the coupling the thermodynamic limit is well defined, and the system reduces to the noisy Kuramoto model for $r=L/2$~\cite{sakaguchi88}.
The fluctuation terms $\xi_{\theta,i}$ and $\xi_{x,i}$ represent two
uncorrelated Gaussian noises such that $\langle
\xi_{\theta,i}(t)\rangle=\langle \xi_{x,i}(t)\rangle=0$, and
$\langle \xi_{\theta,i}(t)\xi_{\theta,j}(t')\rangle=\langle
\xi_{x,i}(t)\xi_{x,j}(t')\rangle=\delta_{i,j}
\delta(t-t')$. The strength of angular fluctuations is determined by
the angular diffusion coefficient $C$, while the spatial
diffusion coefficient $D$ determines the mobility of oscillators. 
The oscillators are point-like particles that can diffuse freely, \emph{i.e.} their movement is not affected by the presence of other oscillators.
Both the phase $\theta_i(t)$ and position $x_i(t)$ are periodic variables such that $0 \leq \theta_i(t) \leq 2 \pi$ and $0
\leq x_i(t) \leq L $. 
\begin{figure}[t]
\centering\resizebox{13.5cm}{!}{\rotatebox{0}{\includegraphics{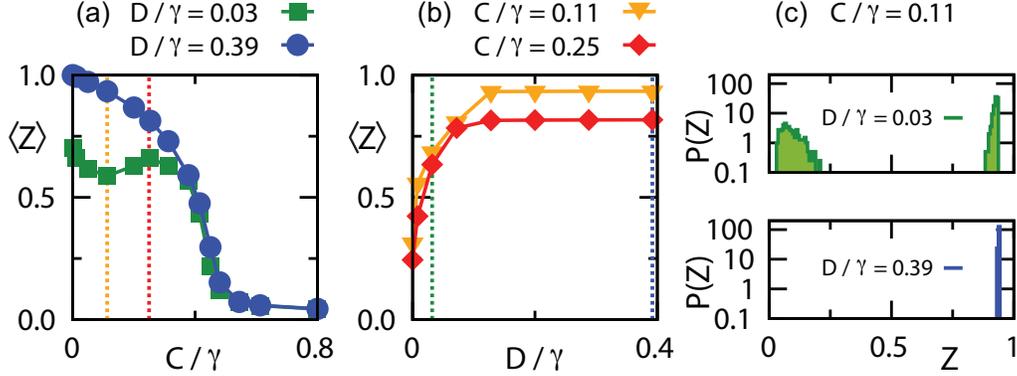}}}
\caption{Twisted states affect the ensemble average of the global order parameter, $\langle Z \rangle$.
(a) $\langle Z \rangle$ vs. scaled angular diffusion $C/\gamma$ for $D/\gamma=0.03$ (green squares)
and $D/\gamma=0.39$ (blue dots).
For small $D/\gamma$, $\langle Z \rangle$ does not reach one, even for $C=0$.
Vertical dotted lines correspond to values of $C/\gamma$ in (b).
(b) $\langle Z \rangle$ vs. scaled mobility $D/\gamma$, for $C/\gamma=0.11$ (orange triangles) and $C/\gamma=0.25$ (red diamonds). Notice that as  $D$ is reduced, $\langle Z \rangle$ decays.
Vertical dotted lines correspond to values of $C/\gamma$ in (a).
(c) The histogram of $Z$ splits into two different peaks at low and high values of $Z$ for low values of $D/\gamma$,
but displays only one peak at high $Z$ for high $D/\gamma$.
Other parameters are $N=1000$, $L=2\pi$, $r/L=0.01$, $\omega=0$ and $\gamma=0.1$.
} \label{fig:R}
\end{figure}

\section{Global order parameter}
The system described by Eqs.~(\ref{eq:osc_i_theta})-(\ref{eq:osc_i_x}) displays a range of states illustrated in Fig.~\ref{fig:snapshots}.
We can characterize global order by the order parameter $Z$,
defined as the absolute value of the population average of the complex unit vectors of the oscillator phases
$Z(t)=N^{-1} |\sum_{j}{e^{i \theta_j(t)}}|$~\cite{kuramoto}.
Mobility introduces a dramatic change in the ensemble average of the order parameter $\langle Z \rangle$, 
where $\langle\dots\rangle$ denotes an average over different initial conditions and realizations of the noise,~Fig.~\ref{fig:R}. 
When mobility is large enough, the system displays global synchronization as $C\to 0$, blue dots in Fig.~\ref{fig:R}(a).
However, when mobility is reduced, global order is compromised, as indicated by the green squares in Fig.~\ref{fig:R}(a).
Increasing mobility $D$ results in an increase of the ensemble average $\langle Z \rangle$, Fig.~\ref{fig:R}(b).
The cause for this behavior can be traced back to the existence of twisted states,
which display local order but have themselves a vanishing global order parameter, Fig.~\ref{fig:snapshots}(c,d).
Due to the presence of these states, $\langle Z \rangle$ results from the average of a bimodal distribution $P(Z)$, with a peak related to the twisted states and the other to global synchronization. 
Above a critical value of the mobility, $P(Z)$ becomes unimodal, Fig.~\ref{fig:R}(c). 
Thus, although a global parameter is not suited to capture the complexity of a system with local interactions, its statistics reflect the existence of twisted states.

\section{Statistical description}
The role of twisted states can be studied using a statistical description that is valid when the number of oscillators is large.
Given that the oscillators have identical autonomous frequencies $\omega$, it is convenient to make the transformation $\theta \to \theta -\omega t$ to a rotating reference frame.
We coarse grain the microscopic model and describe the system in terms of $\rho(x,\theta, t)$, 
the density of oscillators at position $x$ with phase $\theta$, which obeys the Fokker-Planck equation
\begin{eqnarray}\label{eq:FP}
\partial_t \rho(x,\theta,t) &=& D \partial_{xx} \rho(x,\theta,t) + C \partial_{\theta \theta}\rho(x,\theta,t) \\
\nonumber &&+\frac{\gamma}{n(x)} \partial_{\theta} \bigg[ \int_{0}^{L} \! \! \! dx' \! \! \int_0^{2 \pi} \! \! \! \! \! d\theta'  {g(x-x')} \sin(\theta-\theta')  \rho(x',\theta',t) \rho(x,\theta,t) \bigg] \,,
\end{eqnarray}
where $g(x-x')$ is a kernel accounting for the range
and relative strength of local interactions, while $$n(x)=\int_{0}^{L} dx' \int_0^{2 \pi} d\theta' g(x-x') \rho(x',\theta',t)$$ 
denotes the effective number of oscillators in this range. 
In this paper we choose $g(x-x')=1$ for $|x-x'|<r$ and $g(x-x')=0$ otherwise, as in Eq.~(\ref{eq:osc_i_theta}).
The derivation of Eq.~(\ref{eq:FP}) relies on the assumption that $\rho_2(x,\theta,t;x',\theta',t)=\rho(x,\theta,t)\rho(x',\theta',t)$~\cite{hildebrand07}. 

Since the movement of the oscillators is purely diffusive, see Eq.~(\ref{eq:osc_i_x}), 
the spatial density of oscillators is uniform in steady state, $\int_0^{2 \pi} d\theta \rho(x,\theta,t) = N/L \equiv \rho_0$, and $n(x)=2r\rho_0$. 
For small $N$, fluctuations in the spatial density can induce the formation of gaps in which the nearest oscillator is beyond the range of interaction. 
In this paper we consider large densities such that the lifetime of these gaps is much shorter than other typical time-scales.

\subsection{Local order parameter}
The statistical description (\ref{eq:FP}) can be cast in a more transparent form introducing a local mean field.
Local order can be characterized by a local order parameter~\cite{shiogai03,kuramoto06}
\begin{equation}\label{eq:local_kuramoto}
R(x,t)e^{i\psi(x,t)} =\! \!  \int_0^{L} \! \! \! dx'\! \!  \int_0^{2\pi} \! \! \! \! \! d\theta' \frac{g(x-x')}{n(x)} \, e^{i \theta'} \! \! \rho(x',\theta',t) \, ,
\end{equation}
where $R(x,t)$ is a measure of local order and $\psi(x,t)$ is the local average of the phase.
Eq.~(\ref{eq:FP}) can be expressed in terms of this local order parameter as
\begin{eqnarray}\label{eq:FP_localmeanfield}
\partial_t \rho(x,\theta,t) &=& D \partial_{xx} \rho(x,\theta,t) + C \partial_{\theta \theta}\rho(x,\theta,t) \\
\nonumber &+& \gamma R(x,t) \partial_{\theta} \left( \sin \left(
\theta - \psi(x,t) \right) \rho(x,\theta,t) \right) \,,
\end{eqnarray}
reflecting the fact that $\psi(x,t)$ acts as a local mean field and
$R(x,t)$ is a local modulation to the coupling strength.

\section{Transition from disorder to local order}
Eq.~(\ref{eq:FP}) has a trivial steady state $\rho(x,\theta,t)=\rho_0/2\pi\equiv\rho_d$ which corresponds to the
disordered state of the system. We study the stability of $\rho_d$
by inserting $\rho(x,\theta,t)=\rho_d+\epsilon f(x,t) \cos(\ell\theta)$
in Eq.~(\ref{eq:FP}) and keeping terms of order $O(\epsilon)$~\cite{strogatz91}.
Linear stability analysis reveals that the
disordered solution $\rho_d$ becomes unstable for $\ell=1$
when $C<C^{*}$, with
\begin{equation}\label{eq:stab_disordered}
C^{*}=\gamma / 2 \,.
\end{equation}
This threshold is independent of $\rho_0$ and $D$, and determines the
value of $C$ below which local order sets in.
The critical $C^{*}$ given by Eq.~(\ref{eq:stab_disordered}) coincides with the existence~\cite{sakaguchi88} and stability~\cite{strogatz91} threshold displayed by globally coupled noisy oscillators.

\section{Local order solutions}
\begin{figure}[t]
\centering\resizebox{13.5cm}{!}{\rotatebox{0}{\includegraphics{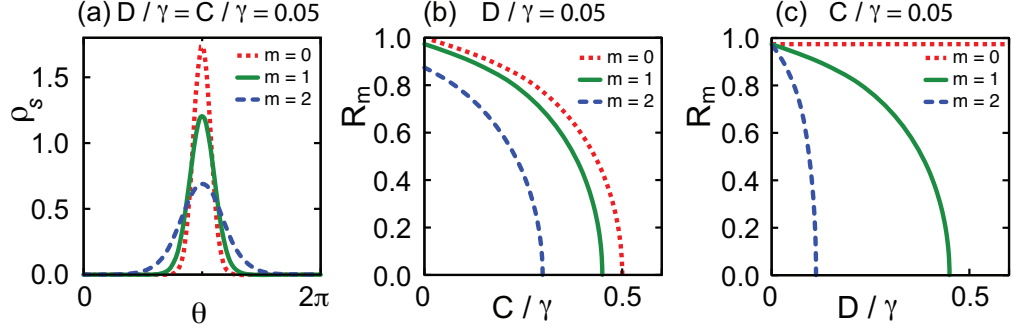}}}
\caption{Twisted solutions and local order.
(a) Angular distribution of steady state twisted solutions, Eq.~(\ref{eq:m-twisted}).
(b) The local order parameter $R_m$ decreases with increasing angular fluctuations for all solutions, Eq.~(\ref{eq:selfconsistency}).
(c) The local order parameter of twisted solutions also decreases with mobility, but is not affected for global order, Eq.~(\ref{eq:selfconsistency}).
The range of interaction is $r/L=0.01$.
} \label{fig:solutions}
\end{figure}

Once local order has set in, the system also supports twisted solutions. 
We specifically look for steady state solutions to Eq.~(\ref{eq:FP_localmeanfield}) of the form $\rho_s(x,\theta)=f(\theta-\psi(x))$.
Such wave-like solutions describe densities in which the angular distribution has the same shape, but is centered at position dependent phases $\psi (x)$.
Setting $\partial_t \rho = 0$ we obtain an ordinary differential equation for $f$ 
\begin{equation}\label{eq:steady_reduced}
\left( D_{\theta} + D_x  \left( \psi'(x) \right)^2 \right) f'(\varphi)  + \gamma R \left( \sin(\varphi) - D_{x} \psi''(x) \right) f(\varphi) = C(x) \,,
\end{equation}
that we can solve together with periodic boundary conditions in phase and space to determine the arbitrary constant $C(x)$ and phases $\psi(x)$.
Periodicity of the phase is consistent with solutions that fulfill $\psi''(x)=0$ and $C(x)=0$, 
and periodicity of space sets the wave numbers $k = 2 \pi m / L$ with $m$ integer.
We obtain the $m$-twist steady state solutions
\begin{equation}\label{eq:m-twisted}
\rho_s(x,\theta) = \norm \exp \left[\frac{\gamma R}{\deff} \cos(\theta - k x ) \right] \, ,
\end{equation}
where $\norm$ is a normalization constant such that $\int_0^{2\pi} d\theta \int_0^L dx \rho_s(x,\theta) = N = L\rho_0$, see Fig.~\ref{fig:solutions}(a).
We have introduced the effective diffusion coefficient
\begin{equation}
\deff (k)= C + k^2 D \, , 
\end{equation}
which is a combination of angular diffusion $C$, and mobility $D$ scaled by the square of the wave number $k$.
Effective diffusion competes with the local coupling $\gamma R$ and controls the width of the angular distribution,
which has a mean $\langle \varphi \rangle = kx$ and variance $$\sigma^2=1-I_1(\gamma R/\deff)/I_0(\gamma R/\deff)=1-R/\sinc(kr),$$
where $\sinc(x)=\sin(x)/x$ and $$I_{n}(z) =\int_0^{2\pi} d\theta \cos^{n}\theta \exp \left( z \cos\theta \right)$$ is the modified $n$-order Bessel function of  the first kind.
The functional form of the steady state angular distribution Eq.~(\ref{eq:m-twisted}) is known as the circular normal distribution~\cite{vonmises18}.

\subsection{Effective diffusion controls the local order parameter}
The $m$-twist solutions described by Eq.~(\ref{eq:m-twisted}) exist if and only if the self-consistency condition obtained by
inserting Eq.~(\ref{eq:m-twisted}) into Eq.~(\ref{eq:local_kuramoto}) is fulfilled
\begin{equation} \label{eq:selfconsistency}
R_m = \sinc(kr) \frac{I_{1}(\gamma R_m/\deff)}{I_{0}(\gamma R_m/\deff)} \, ,
\end{equation}
see Fig.~\ref{fig:solutions}(b,c).
A trivial solution to Eq.~(\ref{eq:selfconsistency}) is $R_m=0$. 
Apart from this, an expansion of Eq.~(\ref{eq:selfconsistency}) for $R_m\ll1$ reveals that non-vanishing solutions
\begin{equation} \label{eq:r_exp}
R_m \approx \sqrt{{8\defc}{\gamma^{-2}}} \left(\defc - \deff \right)^{1/2}
\end{equation}
exist for $\deff \leq (\gamma/2)   \, \sinc(kr) \equiv \defc$.
Again, we see that $\deff$ controls the growth of the local order parameter $R_m$ characterizing the emergence of twisted solutions,
through changes in phase fluctuations $C$ and mobility $D$, Fig.~\ref{fig:solutions}(b,c).

\subsection{Existence of twisted solutions}
\begin{figure}[t]
\centering\resizebox{13.5cm}{!}{\rotatebox{0}{\includegraphics{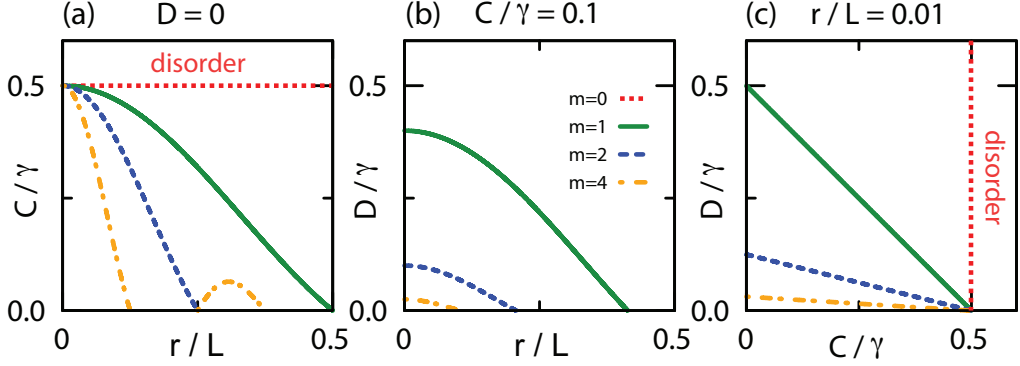}}}
\caption{Mobility controls existence of $m$-twist solutions.
Three representative cuts of the phase diagram are shown for $m=0, 1, 2, 4$. (a) $D=0$, (b) $C/\gamma=0.1$ and (c) $r/L=0.01$.
The dotted red line indicates the onset of local order, which is independent of $D$.
Below the solid green line, the dashed blue line and the dot-dashed yellow line, the $1$-twist, $2$-twist, and $4$-twist solution exists, respectively, see Eq.~(\ref{eq:Dxc}).
An increasing number of co-existing twisted solutions can be found with decreasing mobility $D$ and decreasing coupling range $r$.
} \label{fig:existence}
\end{figure}

We can unfold the effects of spatial and angular fluctuations by writing $\deff$ in terms of its components $D$ and $C$.
Setting $R_m=0$ in Eq.~(\ref{eq:r_exp}) we get
\begin{equation} \label{eq:Dxc}
C + \left({2\pi m}/{L}\right)^2 D = (\gamma/2) \, \sinc \left( {2\pi m r / L} \right)  \, ,
\end{equation}
where we have expressed $k$ in terms of $L$ to stress system size dependence.
For each value of $m$ and range of interaction $r$, the surface in parameter space defined by Eq.~(\ref{eq:Dxc}) encloses the region where the $m$-twist solution exists.
There are two ways in which fluctuations can destroy twisted solutions, by increasing either angular fluctuations or mobility.
Mobility gets amplified for higher order modes as $m^2$, causing twisted solutions to disappear sequentially as mobility increases, Fig.~\ref{fig:existence}.

For $m=0$, Eq.~(\ref{eq:Dxc}) reduces to $C=C^{*}=\gamma/2$ and corresponds to global synchronization.
Existence of global synchronization in steady state is not affected by mobility, as indicated by the dotted red line in Fig.~\ref{fig:existence}(a,c).
However, existence of twisted solutions in finite systems is controlled by angular diffusion $C$, mobility $D$, and range of interaction $r$ as indicated by Eq.~(\ref{eq:Dxc}), Fig.~\ref{fig:existence}.

As $L\to\infty$, the critical value of the spatial diffusion coefficient diverges for all $m$. Therefore, in the infinite system size limit all twisted solutions coexist with global synchronization for any finite $D$. 
These result is in agreement with~\cite{mermin66}, and indicates that identical noisy phase oscillators cannot exhibit a global synchronized state in 1D in this limit.

\subsection{Stability of twisted solutions and states}
\begin{figure}[t]
\centering\resizebox{13.5cm}{!}{\rotatebox{0}{\includegraphics{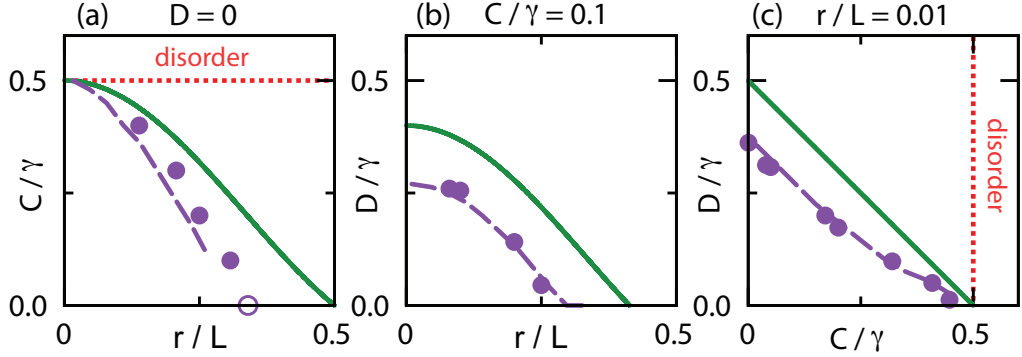}}}
\caption{The domains of existence and stability of $m$-twist solutions do not coincide.
The $1$-twist solution exists below the solid green line,~Eq.~(\ref{eq:Dxc}), but it becomes stable below the dashed purple line, determined numerically.
The purple dots show the stability of the $1$-twist state of the Langevin Eqs.~(\ref{eq:osc_i_theta})-(\ref{eq:osc_i_x}).
The open circle in (a) corresponds to the limit case $D = C = 0$ studied in~\cite{wiley06}.
} \label{fig:stability}
\end{figure}

While existence and stability thresholds coincide for the global order solution, Eqs.~(\ref{eq:stab_disordered}) and~(\ref{eq:Dxc}), this is not the case for $m$-twist solutions in finite systems, Fig.~\ref{fig:stability}.
We address the stability of the $1$-twist solution by performing a numerical study of Eq.~(\ref{eq:FP}), using a finite difference scheme. 
To estimate the stability boundary, we continue a stable twist solution until it becomes unstable against small perturbations. 
We find that the instability is of modulational type.  
The $m$-twist solutions become stable only after the corresponding local order parameter $R_m$ becomes larger than a certain value,
\emph{i.e.} twisted solutions become stable with a finite amplitude, Fig.~{\ref{fig:stability}}.
For vanishing spatial and angular diffusion $C=D=0$, we encounter the system studied by Wiley {\it et al.}~\cite{wiley06}, see purple open circle in Fig.~\ref{fig:stability}(a).
The numerical solution seems to approach this point as $C/\gamma\to0$, but the numerics become lengthy in this limit.

We next compare the continuum Fokker-Planck description~(\ref{eq:FP}) with the discrete system, by means of Langevin simulations of Eqs.~(\ref{eq:osc_i_theta})-(\ref{eq:osc_i_x}).
%
%
To measure the stability of the $1$-Twist state in Langevin simulations, we prepare the system in the $1$-Twist state by randomly positioning the oscillators in space, and setting their phase to $\theta(x)=2\pi x/L$.
We first let the system relax for $1000$ units of time, so that the stationary shape of the angular distribution is reached.
Then, we compute the ensemble average of the twist $\langle m \rangle$, 
where $\langle ... \rangle$ denotes an average over $100$ realizations of the numerical simulation, which we performed for different sets of parameter values $C$, $D$, and $r/L$. 
To evaluate the stability of the $1$-Twist solution we set a threshold on the average twist: when $\langle m \rangle$ falls below $0.05$, we take the $1$-twist state to be unstable. 
Simulations were performed with $N=1000$ and $L=2 \pi$. Other parameters are indicated in Fig.~\ref{fig:stability}.
The stability threshold in Fig.~\ref{fig:stability}(a) was found by varying $r$ for various fixed values of $C$, 
in Fig.~\ref{fig:stability}(b) by varying $D$ for fixed values of $r$, 
and in Fig.~\ref{fig:stability}(c) by varying $C$ for fixed values of  $D$.
We find a good agreement between the stability of the $1$-twist state of the discrete system, 
and the stability of the $1$-twist solution found by numerical integration of the Fokker-Planck equation.

\section{Attraction basins}
Twisted solutions co-exist with global order and among themselves, Fig.~\ref{fig:existence}.
As mobility is increased from low values, twisted solutions become unstable one by one, \emph{e.g.} Fig.~\ref{fig:existence}(c) and Fig.~\ref{fig:stability}(c), 
and global order is enhanced resulting in an increasing value of the ensemble average of the global order parameter as displayed in Fig.~\ref{fig:R}(b).
Angular fluctuations can also destabilize twisted solutions, Fig.~\ref{fig:existence}. 
However, while increasing angular fluctuations can decrease the number of twisted solutions competing with global order, 
the width of the angular distribution corresponding to global order also grows as a consequence of increasing angular fluctuations.
The interplay between these two competing effects may be the cause for the non-monotonic behavior 
observed in the ensemble average of the global order parameter as angular fluctuations $C$ increase, see green squares in Fig.~\ref{fig:R}(a).
For these reasons, it is interesting to explore the size of the attraction basins of the different states that the discrete system exhibits.

\begin{figure}[t]
\centering\resizebox{13.5cm}{!}{\rotatebox{0}{\includegraphics{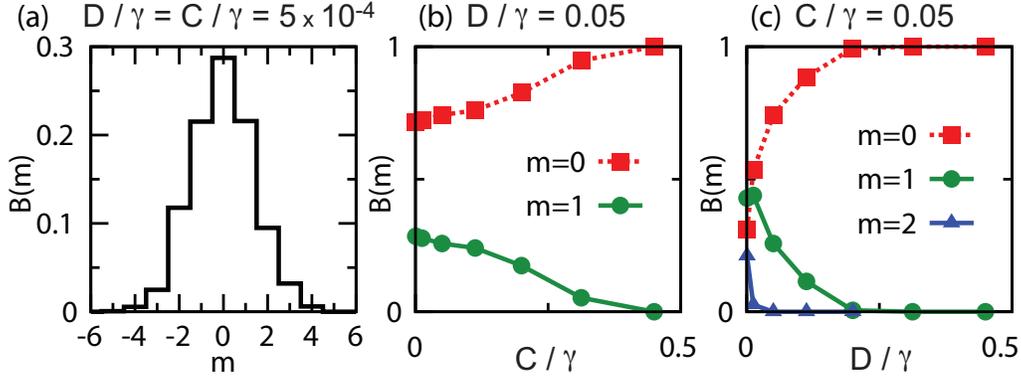}}}
\caption{The size of attraction basins of m-twist solutions depends on angular fluctuations and mobility.
(a) Probability $B(m)$ that starting from a random initial condition the system ends up in a $m$-twist solution.
$B(0)$ and $B(1)$ as function of (b) $C/\gamma$ and (c) $D/\gamma$.
Parameters as in Fig.~\ref{fig:R}.}
\label{fig:basins}
\end{figure}

The fraction of realizations $B(m)$ in which the system ends up in a particular state after a short time, starting from random initial conditions, is a measure of the size of the basins of attraction of the state.
Coexistence of twisted states with up to $m=5$ is shown in Fig.~\ref{fig:basins}(a) for low mobility and phase fluctuations, corresponding to the bottom left corner of Fig.~\ref{fig:existence}(c).
As $C$ is decreased, the attraction basin corresponding to global order shrinks, while the attraction basins of the $m$-twist states expand, Fig.~\ref{fig:basins}(b).
Decreasing $C$ does not necessarily yield global order at the ensemble level because the global ordered state shares the phase space with the $m$-twist solutions:
the observed value of $\langle Z \rangle$ results from the competition between an increase in local order for $m=0$ and a decrease of $B(0)$, see Fig.~\ref{fig:R}(a).
%
%
An increase in the mobility $D$ leads to a contraction of the attraction basin of the $m$-twist states, in favor of global order, Fig.~\ref{fig:basins}(c).
When all $m$-twist solutions are unstable, $B(0)=1$ and the global order state is the only attractor below $C^{*}$, see also Fig.~\ref{fig:R}(b).

\section{Discussion}

We have investigated the effects of mobility in a generic 1D model of locally coupled moving phase oscillators, 
and showed that oscillator mobility dramatically affects the collective behavior of finite systems. 
%
More specifically, our results show that in low dimensional systems global synchronization is compromised by the presence of multiple $m$-twist states exhibiting local order. 
At the onset of local order, the system can fall into the global synchronized state. However, the coexistence of local 
order $m$-twist states implies that the attraction basin of the  global synchronization state is reduced.  
Strong mobility of the oscillators destabilizes these $m$-twist states, and thus promotes global synchronization.

In this paper we have considered a high density limit such that the connectivity of the system is never interrupted by gaps.
%
%
In the dilute limit, gaps in the connectivity play a crucial role in the synchronization dynamics. 
This problem was studied in the context of moving neighborhood networks, and under the assumption of a fast exchange of neighbors, 
a mean-field condition for the existence and stability of the global synchronization state has been derived~\cite{skurfa04}. 
According to this study, whenever the global synchronized state is stable the system reaches global synchronization, regardless of its spatial dimensionality.
In other words, the study overlooks the possibility of coexistence of multiple solutions. 
Our findings reveal a different role for mobility, unrelated to the existence and stability of the global synchronization state: 
mobility disrupts all these multiple solutions except for the global synchronized state. 
Extensions of the current study towards dilute systems will be the subject of further investigations. 

Two-dimensional systems display a similar phenomenology, though the competing local order states can now take other forms, \emph{e.g.} vortexes~\cite{kosterlitz73}. 
%
%
It has been recently reported that chaotic oscillators moving in a two-dimensional space can synchronize provided that spatial dynamics is fast enough~\cite{frasca08}. 
A related albeit different scenario occurs with chaotic advection mixing in two dimensional systems, 
where synchronization of excitable media is enhanced by strong mixing~\cite{solomon04,kurths05}. 
These results indicate that mobility may also enhance synchronization in two-dimensional systems. 
We speculate that global synchronization may be achieved by destabilizing local deffects, as we show here for 1D systems. 
%
%
Further work is intended to clarify these issues.

The theoretical framework introduced here may provide insight into other related problems, as when movement is coupled to the oscillator phases.
In this case synchronization can be interpreted as collective motion~\cite{peruani08}.
As a result of this coupling, strong spatial fluctuations and clustering effects dominate the system dynamics~\cite{zanette04}, 
and global order prevails even in the thermodynamic limit~\cite{toner05}.

Finally, a compelling biological application of our framework may be found in the vertebrate segmentation clock, 
where global coupling is a good effective description of the system because of the high mobility of cells~\cite{riedel07}:
by precluding the appearance of local defects, mobility promotes global synchronization.
Moreover, it has been recently shown that mobility decreases the relaxation times to achieve synchronization in a model 
of the segmentation clock that allows for flipping between neighboring cells~\cite{uriu10}.
However, this system also hosts spatial patterns~\cite{morelli09}, and mobility is not accounted for in current distributed models.
It will be interesting to see how mobility affects the synchrony recovery times and pattern reorganization after perturbation in such models.

%

\subsection*{Acknowledgements}
We thank S. Ares, H. Chat{\'e}, M. Mat\'{\i}as, A. C. Oates and D. Zanette for valuable comments and discussions. 
FP acknowledges financial support from the French ANR projects {\it Morphoscale} and {\it Panurge}. 
LGM acknowledges support from CONICET, ANPCyT PICT 876, and ERC grant 207634 SegClockDyn.



\end{document}